# Eulerian numerical modeling of contaminant transport in Lower Manhattan, New York City, from a point-source release under the dominant wind condition: Insights gained via LES

Wayne R. Oaks[1], Kevin Flora[1], Ali Khosronejad[1*]

[1]Civil Engineering Department, College of Science and Engineering, Stony Brook University, Stony Brook, NY 11794, U.S.A.

*Corresponding author, Email: ali.khosronejad@stonybrook.edu

**Abstract**

Released pollutants and poisonous chemicals in the highly-populated urban area can spread via wind flow and affect the health, safety, and wellbeing of those close to the source and others downstream. Being able to predict the transport of airborne contaminants in such urban areas could help stakeholders to plan emergency responses. Airflow due to wind in urban areas with closely packed buildings results in complicated aerodynamics around buildings that, in turn, can lead to intricate contaminant transport phenomena. We conduct high-resolution large-eddy simulations (LES) of airflow in Southern Manhattan, a highly-populated area in the New York City, due to a dominant wind blowing from south to north. The LES results show complex flow dynamics around the buildings of various heights and, consequently, complicated contaminant transport processes from a source-point which is located in front of the New York Stock Exchange building at 30 m above ground. The flow tends to align with the streets that are fairly parallel to the wind direction creating high-velocity core regions, while the airflow loses its momentum markedly farther downwind in the urban area. For as long as the source of contamination exists, the pollution plume rapidly travels through the streets reaching downwind areas. When the source is removed, contaminants that are trapped in the low-momentum urban area linger to exit the flow domain.

**Keywords:** LES, Eulerian contaminant transport, Micro-scale model, New York City, Biochemical contamination

# 1 Introduction

The transport of contaminants in the form of gas and/or particles through the air could affect the health and wellbeing of people worldwide. In the last two decades, the release of chemical or biological agents has been of great concern after terrorist attacks suffered by the United States in 2001. For an earlier example, on the night of December 2, 1984, a Union Carbide plant in Bhopal, India released toxic methyl isocyanate gas. Since gas is denser than air and was leaked in large quantities, it spread along the ground and into inhabited parts of the city killing at least 3,000 people and injuring about 200,000 (Varma and Guest, 1993). Furthermore, releases of pollutants into the atmosphere affects everyone's health and quality-of-life, as well as the health of the planet. While sulfur dioxide and nitrogen oxide, the precursors to acid rain, have been eliminated countries in the Northern Hemisphere, many other transportation related pollution sources remain unregulated, which led to environmental problems such as acid rain on coastal areas negatively impacting coral reefs (Menz and Seip; 2004, and Risk, 1999). Also, automobiles and factories produce emissions that can affect breathing and lead to lung disease (Sharma et al., 2013). These effects seem to be particularly prominent in crowded urban areas (Jensen et al., 2009; Zhang et al., 2020), where a better understanding of the airborne transport of contaminant could lead to solutions for mitigating such impacts. In the recent decades, Computational Fluid Dynamics (CFD), which is the focus of this study, has been widely used to predict the trajectory and fate of these agents in the air under various wind conditions. The CFD-based studies seek to generate reliable information about the transport of airborne contaminants to help decision makers and stakeholders mitigate the hazardous impact of airborne contaminants in communities (Pontiggia et al., 2010).

Over the past several decades, micro-scale numerical models, (i.e., those that simulate regions of 2 to 20 km) have been utilized to investigate the transport of contaminants around buildings in populated urban areas. Single obstacles or buildings have been studied to find the applicability of CFD for predicting flow structures and patterns of pollutant transport (Murakami, 1993; Rossi et al., 2010). In studies at the beginning of the 21$^{st}$ century, single obstacle models used coarse meshes because of the limited computing power of the processing machines (Tominaga and Stathopoulos, 2013). Additionally, there were no best practice guidelines at the time, however, more recent studies have benefited from the advancement in computational power and best practice guidelines. Two dimensional (2D) and three dimensional (3D) numerical studies of buildings with varying height, width, and length have revealed the existence of complex flow patterns around individual buildings (Aristodemou et al., 2020). Past numerical studies have also attempted to predict wind flow patterns and plume dispersion around real-life building complexes (Gousseau et al., 2011). For example, Guo et al. (2020) conducted numerical modeling of arrays of regularly spaced objects using a Reynolds Averaged Navier-Stokes (RANS) based model. They found that the flow aligned perpendicular to the front array objects contained the most pollution in the first canyon which tended to stay there. In contrast the oblique flow below the tops of the objects follows the channels and the more elevated flow follows the wind direction. Santiago et al. (2010) studies arrays of regularly spaced objects with geometrical irregularities and using large-eddy simulation (LES) model.

The next level of numerical modeling efforts studied realistic city geometries. The main concern for urban street canyons has been the distribution of pollutants (Gallagher and Lago, 2019; Wang, 2008). Past studies have shown that if wind flows transverse to and on the top of the canyon it causes recirculation zones that flow down the windward wall and up the leeward wall. Furthermore, Wang (2008), and Gallagher and Lago (2019) show that pollution concentrations are higher at the street level and reduces as the elevation increases. They also showed that pollution concentration is relatively higher in the leeward side of the canyon and that obstacles, such as parked cars along the roads tend to reduce the pollution concentration on sidewalks. Overall, it has been shown that as the number of realistically modeled buildings increases, the resulting flow fields become more complex (Hanna et al., 2006; Gousseau et al., 2011; and Aristodemou et al., 2020). The simulation results of the past numerical studies have been validated using experimental data of small-scale models of partial sections of cities such as Montreal (Gousseau et al., 2011), Oklahoma City (Neophytou et al., 2011), Tokyo (Ashie and Kono, 2011), and New York City (Hanna et al., 2006; and Tominaga and Stathopoulos, 2013).

As mentioned above, past numerical studies have contributed a great deal to our current understand about wind flow and pollution transport in urban areas (Zheng et al., 2010; Nozu and Tamura, 2012; Sanchez et al., 2017; Toja-Silva et al., 2017; Schatzmann and Leitl, 2011; Harms et al., 2011; and Tominaga and Stathopoulos, 2013). To the best of our knowledge, nearly all of these studies use body-fitted computational grid system. In this study, we employ our in-house model, the so-called Virtual Flow Simulator (VFS-Geophysics) mode, which has several key characteristics that makes it more relevant for micro-scale modeling of wind flow and pollution transport in urban areas with complex geometry. More specifically, the VFS-Geophysics model uses the immersed boundary method (IBM) which makes it capable of modeling the complex details of building geometries. The turbulent flow of the wind is modeled using the LES model while the Eulerian transport of contaminants is computed using a convection-diffusion model.

Herein, we report a micro-scale model of a portion of the New York City in Lower Manhattan. The detailed geometry of the Lower Manhattan was obtained from LiDAR data of the New York City that is publicly available (https://www.openstreetmap.org; https://www1.nyc.gov/site/doitt/initiatives/3d-building.page). In the context of IBM, the geometry of the city including the roads and buildings were discretized using an unstructured triangular grid system and immersed into the flow domain, which included a 2.5 km long, 1.8 km wide, and 600 m high region of the Lower Manhattan, see Figure 7. We numerically investigated the transport of point-source contaminant released from the east side of the New York Stock Exchange building at approximately 30 m above street level. The wind that blows south to north with a mean-flow velocity of 3.58 m/s under neutral thermal conditions.

In this study, we seek to carry out the LES of the wind flow through lower Manhattan, where skyscrapers are the dominant buildings, coupled with the Eulerian convection-diffusion model of a point-source contaminant transport on a high-resolution computational grid system with over 76 million grid nodes. The point source contaminant studied here resembles the pollution source that can be created as a result of a biochemical bomb. The objective of the study is to gain insights into the mechanism that contribute to the spreading of such contaminants in high-populated areas like New York City. The high-resolution LES in this work enables us to (1) determine the shape and geometry of the contaminant plume spreading throughout the city, (2) determine the hot spots of the contaminant, where contaminants are trapped, (3) find out the time required for contaminant to reach various points of the city, and (4) quantify the resident time of contaminant concentration as the contaminants are gradually advected out of the city.

This paper is organized in the following manner. In Section 2 we present the governing equations of the numerical model. Numerical model validation study is presented in Section 3, followed by the computational details of the New York City model in Section 4. Subsequently, the simulation results are presented and discussed in Section 5. Finally, in Section 6 we conclude the findings of this study.

## 2 Numerical model

The VFS-Geophysics model solves the incompressible form of the Continuity and Navier-Stokes equations and Eulerian convection-diffusion equations to obtain the wind flow and contaminant concentration fields. The equations are solved in their nonorthogonal, generalized, curvilinear coordinates form and over structured grid systems, which we denote as the background mesh. Using the IBM, the arbitrarily complex geometry of the buildings and cityscape are discretized using unstructured triangular grid system and immersed into the background mesh. Starting from the Continuity and Navier-Stokes equations in Cartesian coordinates $\{x_i\}$, both the independent variables and spatial coordinates are transformed to obtain the governing equations in nonorthogonal, generalized, curvilinear coordinates $\{\xi_i\}$, which in the compact tensor notation read as follows (free and dummy indices as integer values of 1 to 3):

$$J \frac{\partial U^i}{\partial \xi^i} = 0 \tag{1}$$

$$\frac{1}{J}\frac{\partial U^i}{\partial t} = \frac{\xi_l^i}{J}\left[-\frac{\partial}{\partial \xi^j}(U^j u_i) + \frac{1}{\rho}\frac{\partial}{\partial \xi^j}\left(\mu \frac{g^{jk}}{J}\frac{\partial u_i}{\partial \xi^k}\right) - \frac{1}{\rho}\frac{\partial}{\partial \xi^j}\left(\frac{\xi_i^j p}{J}\right) - \frac{1}{\rho}\frac{\partial \tau_{ij}}{\partial \xi^j}\right] \tag{2}$$

where $J$ is the Jacobian of the geometric transformation: $J = |\partial(\xi^1, \xi^2, \xi^3)/\partial(x_1, x_2, x_3)|$, $\xi_j^i = \partial \xi^i/\partial x_j$ are the transformation metrics, $u_i$ is the filtered $i^{\text{th}}$ Cartesian velocity component, $U^i = (\xi_m^i/J)u_m$ is the filtered contravariant volume flux, $g^{jk} = \xi_l^j \xi_l^k$ are the components of the contravariant metric tensor, $p$ is the pressure, $\rho$ is the density, $\mu$ is the dynamic viscosity, and $\tau_{ij}$ is the sub-grid stress (SGS) tensor for the LES method (Khosronejad et al., 2016). Instantaneous velocity components are decomposed into the resolved turbulence, calculated directly from the Navier-Stokes equations, and unresolved turbulence that is modeled as the SGS terms. The velocities in the Navier-Stokes equations are replaced with the sum of the resolved and unresolved velocity components and integrated using a spatial filter with SGS terms appearing in the momentum equations. Herein, the SGS terms are closed using the dynamic Smagorinsky SGS model (Germano et al., 1991). The SGS tensor, $\tau_{ij}$, is obtained as:

$$\tau_{ij} - \frac{1}{3}\tau_{kk}\delta_{ij} = -2\mu_t \overline{S_{ij}} \tag{3}$$

where $\delta_{ij}$ is the Kronecker delta, $\mu_t = C_s \Delta^2 |\bar{S}|$ is the dynamic eddy viscosity, $\overline{S_{ij}} = 1/2 \left( \partial u_i / \partial x_j + \partial u_j / \partial x_i \right)$ is the filtered strain rate tensor, $C_s$ is the Smagorinsky constant, $\Delta = J^{-1/3}$ is the box filter size, and $|\bar{S}| = \left( 2\overline{S_{ij}S_{ij}} \right)^{1/2}$ (Khosronejad et al., 2016).

The continuity and momentum equations are solved on the structured background mesh where the governing equations are discretized in space. A hybrid staggered/non-staggered arrangement is used where variables are stored at the cell center, while mass fluxes are stored at the faces (Gilmanov and Sotiropoulos, 2005; Ge and Sotiropoulos, 2007). The second-order accurate central differencing scheme is used to discretize the convective terms, while the divergence, pressure gradient, and viscous like terms are discretized using the second-order accurate, three-point central differencing (Gilmanov and Sotiropoulos, 2005). Furthermore, the second-order backward-differencing scheme is utilized to discretize the time derivatives (Kang et al., 2011). A second-order accurate fraction-step method is used to incorporate a pressure solution into the procedure for solving the momentum equations. The Jacobian free, Newton-Krylov solver is employed to solve the momentum equations. The generalized minimal residual (GMRES) solver, using the multigrid method as a preconditioner, is applied to solve the Poisson pressure equation (Ge and Sotiropoulos, 2007; Borazjani et al., 2008; and Kang et al., 2011).

The arbitrarily complex geometry of the buildings and other cityscape's features were handled using the IBM, in which the geometries are discretized using unstructured triangulated surfaces in 3D space and immersed into the structured background mesh. The governing equations are solved on the background mesh while the velocity field at the nodes adjacent to the solid surfaces of the immersed bodies are reconstructed using the sharp-interface IBM (Borazjani et al., 2008) and a wall model (Khosronejad and Sotiropoulos, 2014) and used as the boundary conditions. The computational nodes of the background mesh that are located inside the immersed bodies are blanked out of the computations (Kang et al., 2011).

The contaminant concentration is treated as a passive tracer and its dispersal is modeled using Eulerian convection-diffusion. As with the flow equations, the Cartesian form of the equation's independent and spatial variables are transformed to obtain the nonorthogonal and curvilinear form of the convection-diffusion equation, as follows (free and dummy indices as integer values of 1 to 3):

$$\frac{1}{J}\frac{\partial \psi}{\partial t} + \frac{\partial}{\partial \xi^j}(\psi U^j) = \frac{\partial}{\partial \xi^j}\left[\left(\frac{v}{S_l} + \frac{v_t}{S_t}\right)\frac{g^{jk}}{J}\frac{\partial \psi}{\partial \xi^k}\right] \quad (4)$$

where $\psi$ is the filtered concentration, $S_l$ is the laminar Schmidt number (=700), $S_t$ is the turbulent Schmidt number (=0.75), $v$ is the fluid kinematic viscosity, and $v_t$ is the eddy kinematic viscosity (Khosronejad and Sotiropoulos, 2018; and Khosronejad and Sotiropoulos, 2020). The convection-diffusion equation (equation 4) is discretized using the second-order central differencing numerical scheme and solved . The discretized form of the convection-diffusion equation is then solved using the fully implicit Jacobian free Newton method (Chou and Fringer, 2008; Chou and Fringer, 2010; Kraft et al., 2011; and Khosronejad et al., 2016).

## 3 Model validation

The model validation study was carried out using the measured data obtained during a series of experimental tests in a wind tunnel at the Saint Anthony Falls Laboratory of the University of Minnesota. The experimental wind tunnel is 1.5 m wide, 1.7 m height, and 16 m long. It can produce air flows with a maximum mean-flow velocity of 45 m/s (see Fig. 1) and is equipped with particle image velocimetry (PIV) to measure the air velocity field.

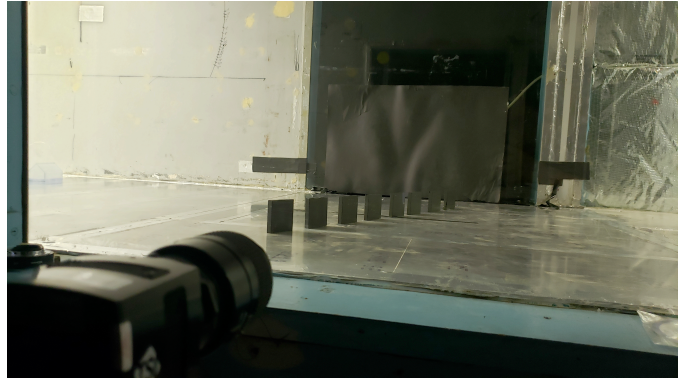

**Figure 1.** An image of the wind tunnel facility at the St Anthony Falls Laboratory of University of Minnesota, where a series of experimental tests were conducted to measure the flow field around 1/50[th] physical model of realistic buildings. Flow is from left to right.

A 1/50[th] physical model of 12 buildings was created and placed on a turn table in the wind tunnel. The physical model resembles a real-life test site at the United States Army Aberdeen Proving Ground in Edgewood Area, Maryland (see Fig. 2). As seen, the testing site is about 100 m wide and 100 m long. The individual buildings are of various size and their height vary between 2.5 m and 3.8 m. The 1/50[th] physical model of the testing site was exposed to air flows with a mean-flow velocity of 3 m/s in the wind tunnel. To test various wind directions, the turn table was placed in three positions to resemble three different wind directions (see Fig. 3). Using the PIV technique, the velocity field was measured at a test cross-plane around the building number four (see Fig. 3). Using the PIV data, we extracted vertical profiles of streamwise velocity component at six local points (three upstream and the other three downstream of the building number four in Fig. 3). The so-obtained PIV data was used to validate the flow solver of the VFS-Geophysics model.

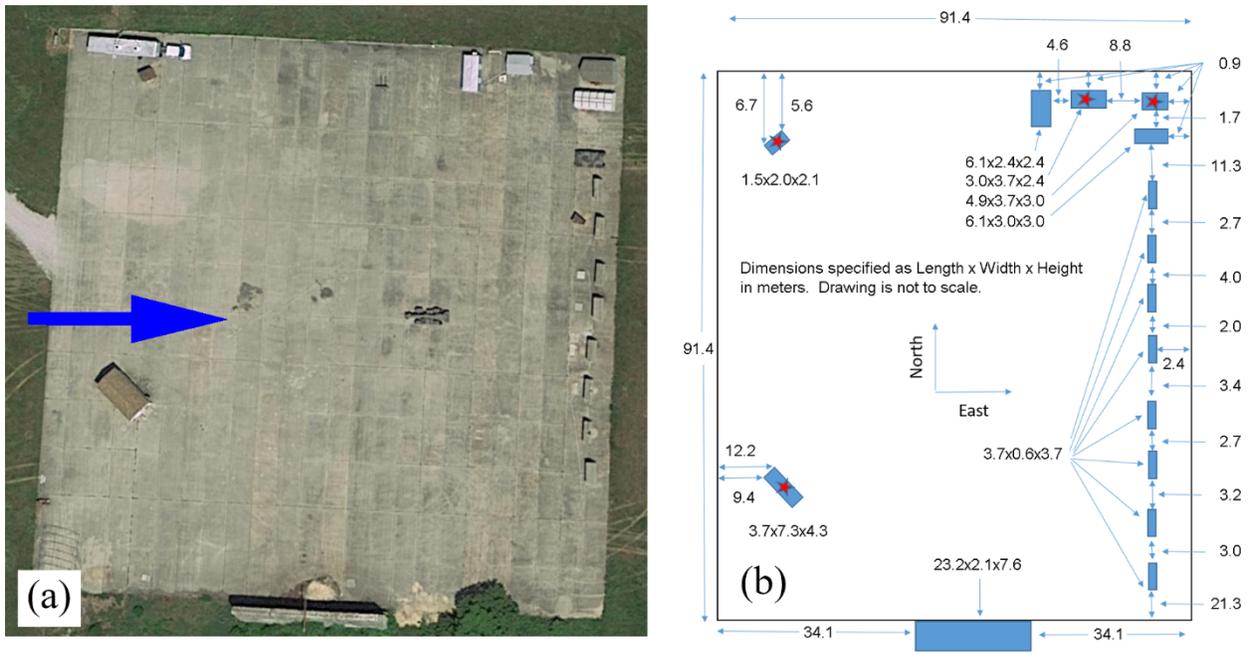

**Figure 2.** Field test site at United States Army Aberdeen Proving Grounds from top view. (a) a real image of the field testing-site. (b) schematic of the field testing-site showing the dimension of building and their relative location. The blue arrows show the direction of flow which is from left (west) to the right (east).

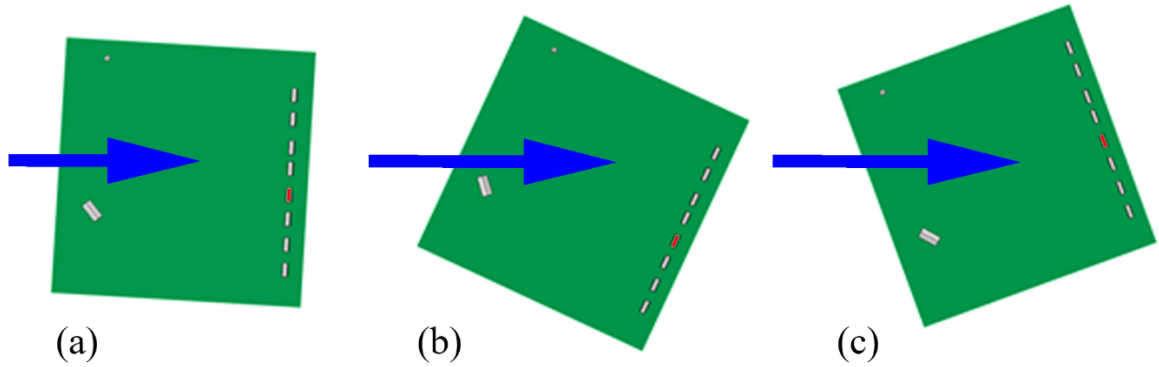

**Figure 1.** The 1/50[th] physical model is placed on a turn table to enable its rotation and generate three different flow directions relative to the axis of the physical model, which construct three configurations. (a) shows configuration 1 with a flow direction from the west at an angle of 3°. (b) presents configuration 2 with a flow direction from south west at an angle of 25°. (c) marks configuration 3 with a flow direction from north west at an angle of 20°. The blue arrows show the direction of flow. The measurement plane intersects the red buildings.

We built the numerical model of the wind tunnel including the three physical models of the buildings. A background mesh consisting of over 168.9 million computational grid nodes was created to contain the wind tunnel. The background mesh covered 4.1 m of the wind tunnel length, where the physical model of the buildings is located. The background mesh extended 1.5 m upstream and downstream of the buildings. A grid resolution of the background mesh was about 3 mm in all directions. A temporal time step of 0.002 s was selected to achieve a Courant Friedrichs Lewy (CFL) number of 1. Moreover, the solid surfaces of the buildings and wind tunnel walls were discretized using unstructured triangular elements and immersed in the background mesh.

To produce a fully turbulent inlet velocity filed as the inlet boundary conditions at the inlet of the numerical model of the wind tunnel, we conducted a precursor LES using a 1.5 m length of the wind tunnel, which had the same grid resolution as that of the background mesh for the entire wind tunnel. Using a periodic boundary conditions in streamwise directions, the precursor LES was performed until the flow field solution converged, i.e., when the total kinetic energy of the flow plateaus. The fully converged LES solution was continued for 5000 another time steps during which the instantaneous flow field was recorded over a cross plane in the mid-length of precursor domain. More details regarding our precursor LES methodology has been extensively documented (Khosronejad et al., 2020; Khosronejad et al.; 2020a; Khosronejad et al., 2020b).

Using the recorded instantaneous turbulent flow of the precursor LES as the inflow boundary conditions, we carried out three simulations for the three configurations. For each configuration, we computed instantaneous flow field of the wind tunnel with the 1/50$^{th}$ physical model of the buildings. The simulations were continued for 1.4 s of physical time until the flow field was statistically converged. As this point, we continued the simulations and time averaged the computed flow fields for about 4.0 s when the time-averaged statistics converged (for details of the convergence criteria, see (Khosronejad et al., 2020; Khosronejad et al., 2020b; Khosronejad et al., 2020b). The time-averaged data for the streamwise velocity component of the three configurations are compared with those of the experimental test in the wind tunnel. As seen in Figs. 4 to 6, the LES computed streamwise velocity for configurations 1 to 3, respectively, agree well with the measured data.

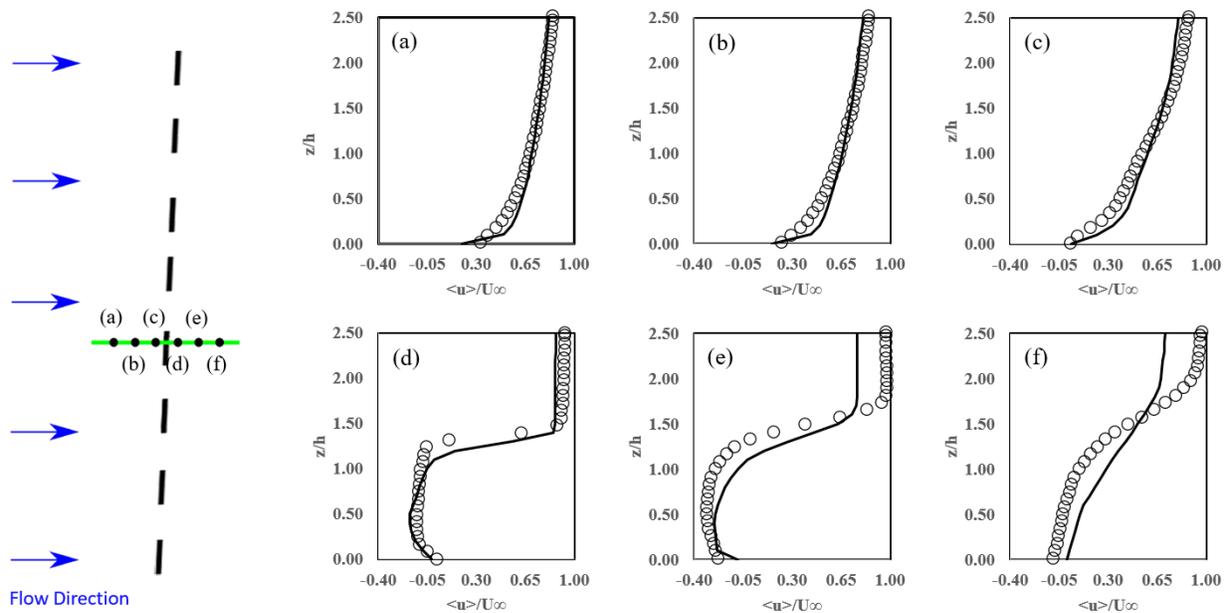

**Figure 2.** Measured (circles) and LES computed (solid lines) time-averaged streamwise velocity for configuration 1. Scheme on the left column plots the relative location of the six local points, where velocity profiles (in vertical) are extracted.

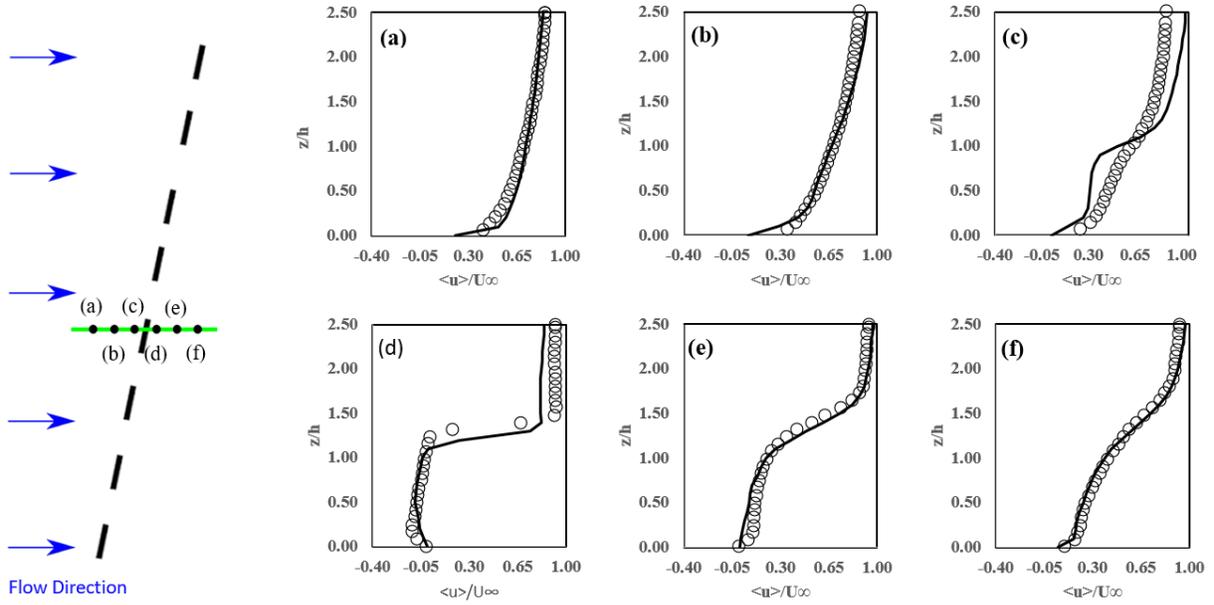

**Figure 5.** Measured (circles) and LES computed (solid lines) time-averaged streamwise velocity for configuration 2. Scheme on the left column plots the relative location of the six local points, where velocity profiles (in vertical) are extracted.

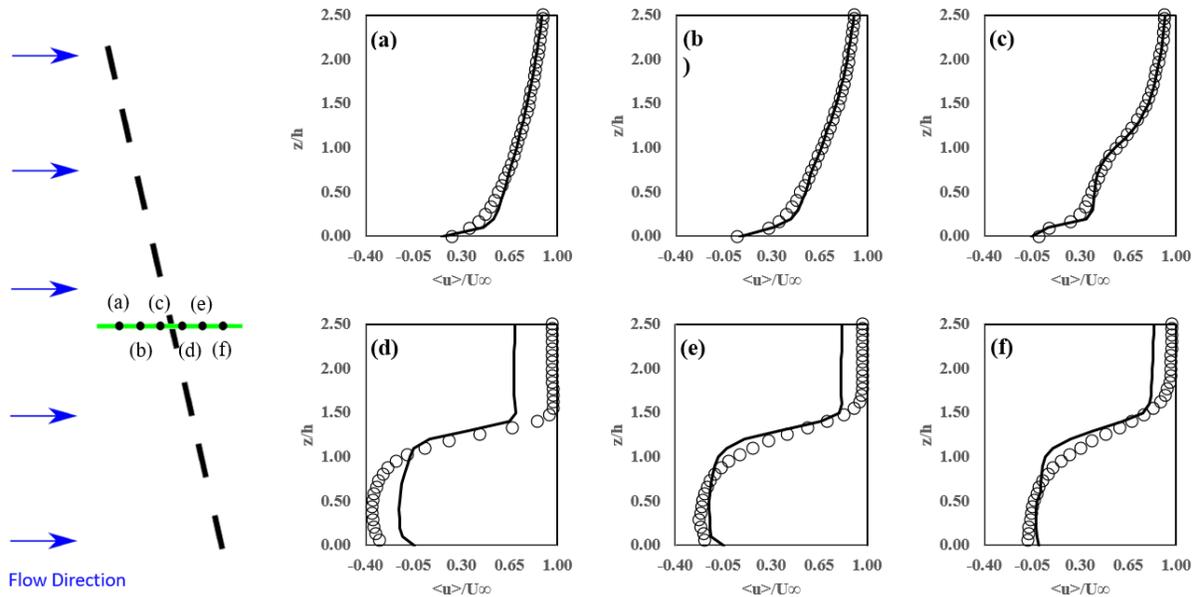

**Figure 6.** Measured (circles) and LES computed (solid lines) time-averaged streamwise velocity for configuration 3. Scheme on the left column plots the relative location of the six local points, where velocity profiles (in vertical) are extracted.

# 4  Computational details

This study is focused on high-fidelity numerical modeling of wind flow and contaminant transport in Lower Manhattan in New York City. The study area and, therefore, the computational domain includes the southern tip of Manhattan, north of Upper Bay, east of the Hudson River, and west of the East River. As seen in Fig. 7, it extends approximately 2.5 km from the southern tip of the island and is bounded by the following roads from east to west: access roads to the Brooklyn Bridge, Park Row, Worth Street., Baxter Street, Canal Street, and Laight Street. Therefore, the computational domain is 2.5 km long, 1.8 km wide, and 0.6 km high. The height of the background mesh is set equal to 0.6 km to be well above height of the tallest building (i.e., the One World Trade Center) in the study area. The background grid system was created so that it encompasses the above-mentioned computational domain of the Lower Manhattan. As shown in Table 1, the background mesh has a grid resolution of 3 m in all directions resulting in a computational grid system with over 76.5 million nodes. The time step of 0.28 s was selected to limit the CFL number to be less than 1. We note that solving a LES with a flow domain involving a large high number of computational nodes can be computationally overwhelming. Tominaga et al., 2008 suggested that the grid resolution for a micro-scale model of an urban area to be about 10% of the smallest building size and ten computational cells to cover the space between each two buildings. Our background mesh meets the former suggestion and mostly meets the latter -- except for some narrow streets.

   We obtained the 3D geometric data of the New York City from open source sites (https://www.openstreetmap.org; https://www1.nyc.gov/site/doitt/initiatives/3d-building.page), which includes the geometry of buildings and grounds, from an official 2014 aerial survey. Using the open-source software programs blender (https://www.blender.org) and meshmixer (http://www.meshmixer.com), we were able to convert the arial survey data of the city to unstructured triangular grid system. The obtained geometry of the city which included more than 364 buildings was then immersed into the background grid system of the study area (see Figure 8) to create a complete digital map of the Lower Manhattan to conduct high-fidelity LES of the wind flow and contaminant transport.

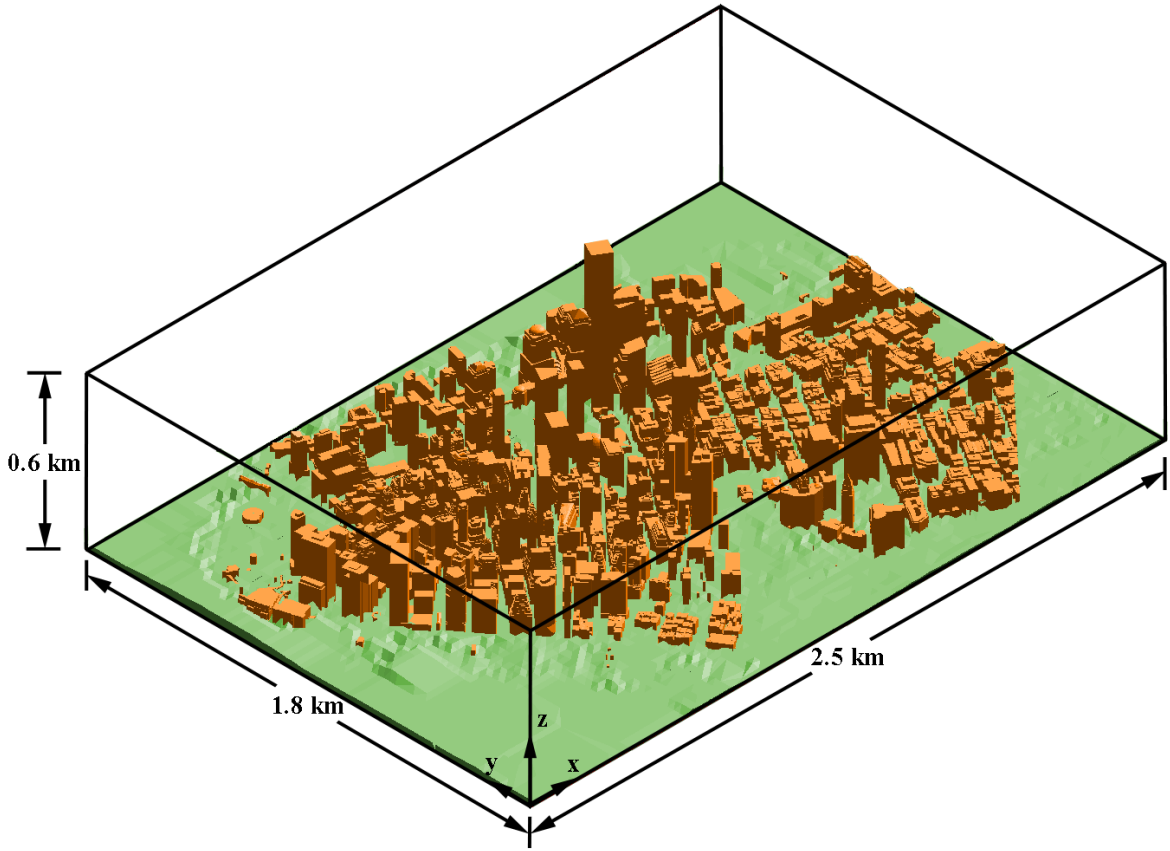

**Figure 7.** 3D view of the study area in New York City, which is limited to Lower Manhattan. The computational domain is 2.5 km long (in the streamwise direction (x)), 1.8 km long (in the spanwise direction (y)), 0.6 km high (in the vertical direction (z)).

Using the Iowa Environmental Mesonet (https://mesonet.agron.iastate.edu) wind rose of 2019, we selected the dominant wind condition in Lower Manhattan to have a mean-flow velocity of 3.58 m/s which blows from south to north. The selected dominant wind conditions had the maximum percentage of occurrence in 2019.

The southern face of the study area was set as the inlet cross-plane of the flow domain while the opposite cross-section (i.e., northern face of the domain) was set as the outlet. The western, eastern, and top faces of the study area are treated using the free-slip boundary condition. A precursor LES was conducted to generate instantaneous turbulent boundary condition (with the mean-flow velocity of 3.58 m/s) to impose as the inlet boundary condition for the main study area (i.e., Lower Manhattan). The procedure for this precursor LES was similar to that we discussed in Section 3. The LES of the wind flow in the study area was continued for enough time until the wind flow throughout the domain reached a dynamics steady state indicated by a steady kinetic energy. At this point, the source-point contaminant concertation was released and allowed to spread. Finally, it should be noted that the simulation was carried out on a supercomputing cluster using 224 CPUs each with a 2.0 GH processor for 2 months of clock-time.

**Table 1.** Computational details of the background grid system of the Lower Manhattan. $N_x$, $N_y$, and $N_z$ are the number of computational grid nodes in streamwise, spanwise, and vertical directions, respectively. $\Delta x$, $\Delta y$, and $\Delta z$ are the special resolution in streamwise, spanwise, and vertical directions, respectively. $z^+$ is the vertical resolution in wall units. $\Delta t$ is the time step.

|  | Background mesh |
| --- | --- |
| $N_x \times N_y \times N_z$ | $833 \times 585 \times 157$ |
| $\Delta x$ (m) | 3.0 |
| $\Delta y$ (m) | 3.0 |
| $\Delta z$ (m) | 3.0 |
| $z^+$ | 24000 |
| $\Delta t$ (s) | 0.28 |

The initial boundary condition for the contaminant concentration field includes a virtual sphere of gas contaminant with a diameter of 24 m and centered at 30 m above the ground. The contaminant source is design so to resemble a biochemical explosion and is located on the east side of the New York Stock Exchange (see Fig. 9). The initial boundary condition is imposed on the computational cells within the sphere of contaminant. The contaminant concentration at these nodes is set to 1.0 (in volume fraction) t = 0. The contaminant concertation was then allowed to spread throughout the study area while keeping the point-source concentration at 1.0. We continued the constant release of contaminant for 46 min, when the point-source contamination was removed. After removing the source of contamination, it took a while for the contaminant to almost exit the study area completely. Details of the contaminant transport during these faces are described in Section 5.

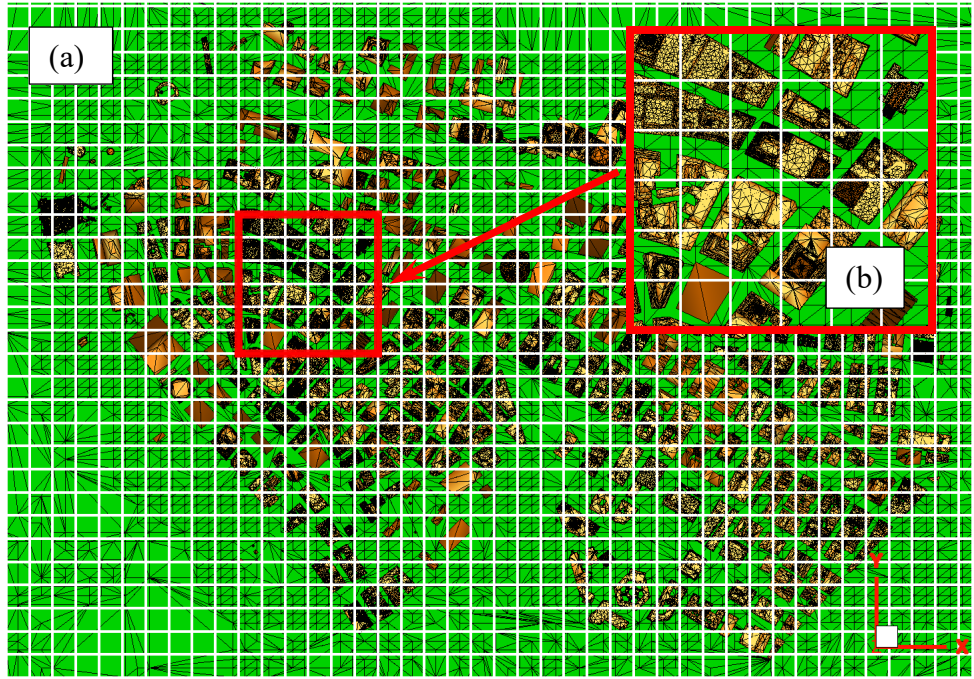

**Figure 8.** Schematic of the proposed immersed boundary approach in the Lower Manhattan from top view. The buildings, road, and other cityscape's features are discretized with an unstructured triangular mesh (in black) and treated as a sharp-interface immersed body embedded in the flow domain discretized with a structured Cartesian mesh (white lines). The nodes of the structured Cartesian mesh located in the buildings (brown), roads and ground (in green) are blanked out of the simulation. Every other 20 background grids are shown in (a) and (b). The enlarged region in (b) shows a zoomed in section centered on the gas contaminant release location.

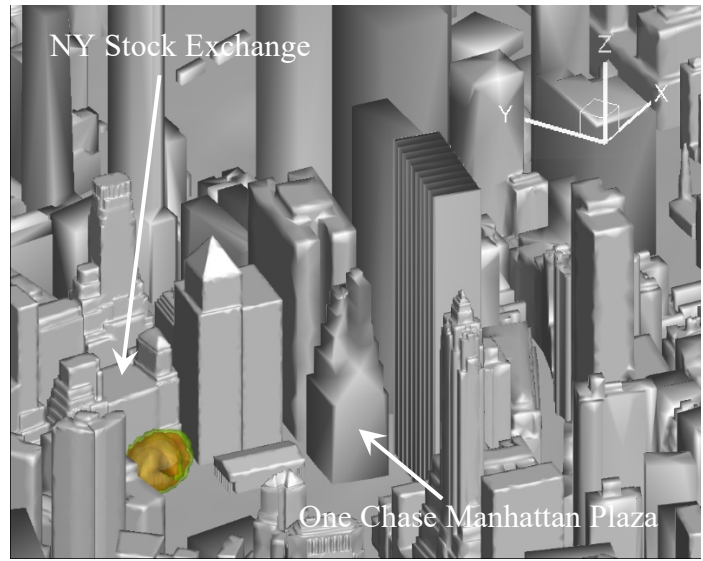

**Figure 9.** 3D geometry of a portion of the study area, where the virtual source of gas contaminations is located. The sphere of contamination source is located east of the New York Stock Exchange building and south of One Chase Manhattan Plaza. The simulated dominant wind flows from south (bottom right) to the north (upper left).

## 5  Results and discussion

In this section, we present the LES results of the wind flow field and contaminant transport from a source-point released in Lower Manhattan and discuss the importance of our findings in terms of the impacts of point-source contaminations on public health in the most populated urban area in North America.

The simulation results are shown over a horizontal plane and three cross-planes in Lower Manhattan, as shown in Fig. 10. As seen in this figure, the cross plane "P1" is in the streamwise direction and passes through the contaminant release source-point. The cross plane "P2" is parallel to New Street, which is located between the New York Stock Exchange building and 86 Broadway. The cross plane "P3" follows the road grid from the contaminant release source-point. Moreover, the horizontal plane on which the simulation results are plotted is located at an elevation of 13.4 m. And, since the ground, in the study area, is not flat, the horizontal plan is not at a consistent distance from the ground. Finally, the buildings in the foreground have been removed to allow for a better view of the simulation results.

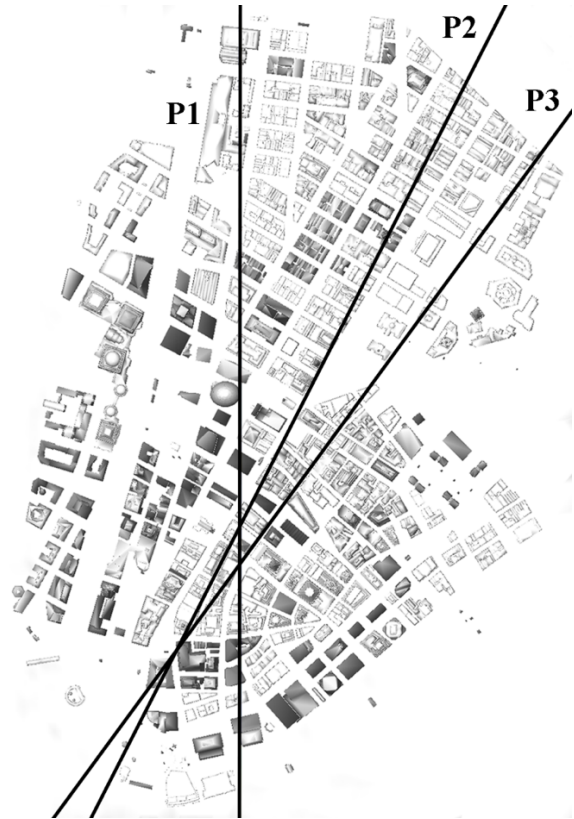

**Figure 10.** The study area in the Lowe Manhattan, New York City, from top view. The black lines "P1", "P2", and "P3" show the vertical cross-planes on which the simulation results are presented. The wind flow is from the south (bottom) to the north (top).

## 5.1 Instantaneous flow field

We start with presenting the instantaneous simulation results of the flow field in the study area. In Fig. 11, we plot contours of LES-computed instantaneous velocity magnitude (normalized with the mean-flow velocity of the dominant wind, i.e., 3.58 m/s) on the cross planes "P1" to "P3" from the side view. As seen, the urban canopy, which includes buildings of various height, act as a large-scale roughness creating low momentum regions among the buildings. The low momentum region is separated from the atmospheric boundary layer by a strong turbulent shear-layer immediately above the urban canopy. In addition, the individual buildings that are taller than the mean height buildings shield the downstream buildings creating a complicated flow field.

A similar flow pattern can be seen in Fig. 12, which shows the contours of LES-computed instantaneous normalized velocity magnitude on a horizontal plane that is approximately 13.4 m above the mean ground level. As seen in this figure, the wind flow is channelized into the streets that are aligned with the wind direction. The wind speed in such aligned streets and alleys can increase significantly and become many times greater than the wind velocity in streets and alleys that are in the wake of buildings (e.g. see the high wind velocity of West Street in Fig. 12). The wind velocity in the aligned streets and alleys, however, decreases quickly as the flow enters deeper into the interior of the city, where the low momentum areas, caused by the urban canopy are abundant.

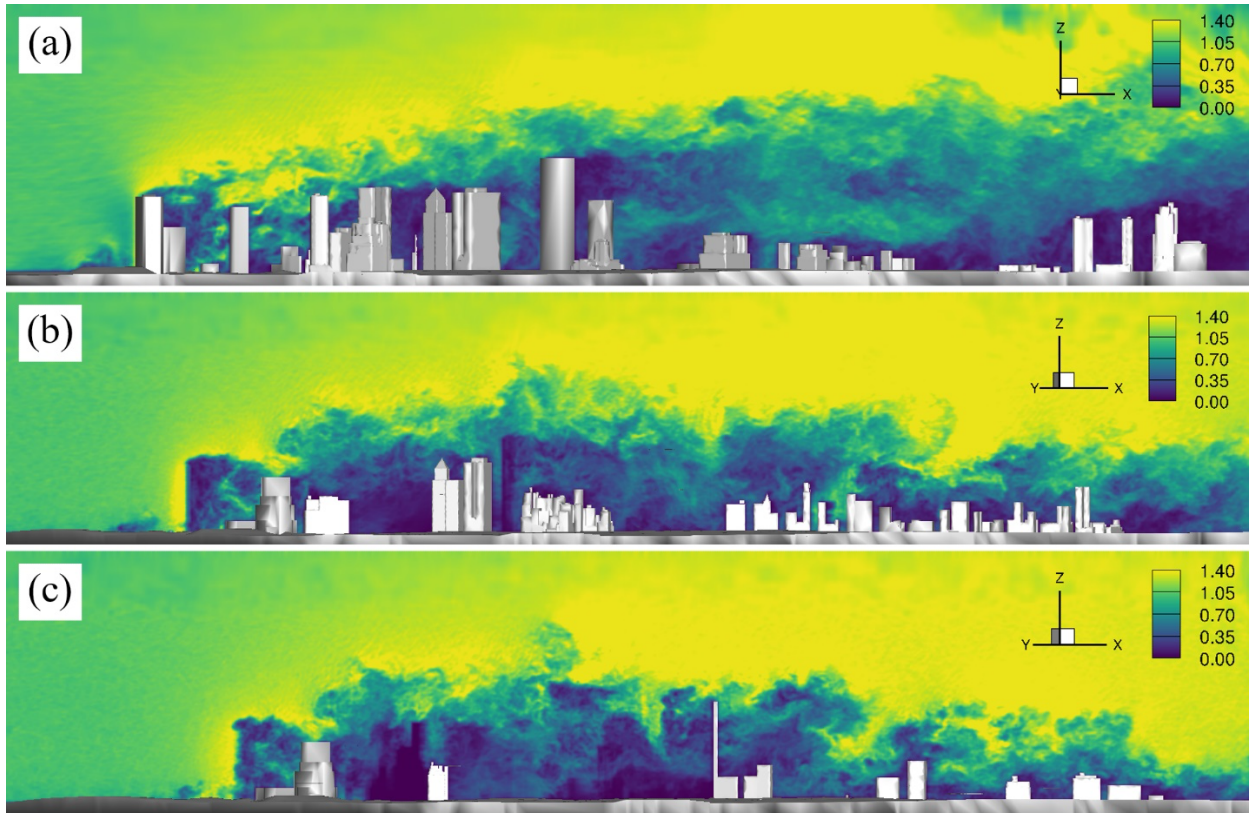

**Figure 11.** Contours of LES-computed instantaneous velocity magnitude, non-dimensionalized with the mean-flow velocity of the dominant wind (=3.58 m/s), on the cross planes "P1" (a), "P2" (b), and "P3" (b). Location of the cross planes is shown in Fig. 10. Buildings and ground are shown in grey. The tallest building in (a) is the One World Trade Center. The wind flow is from the south (left) to the north (right).

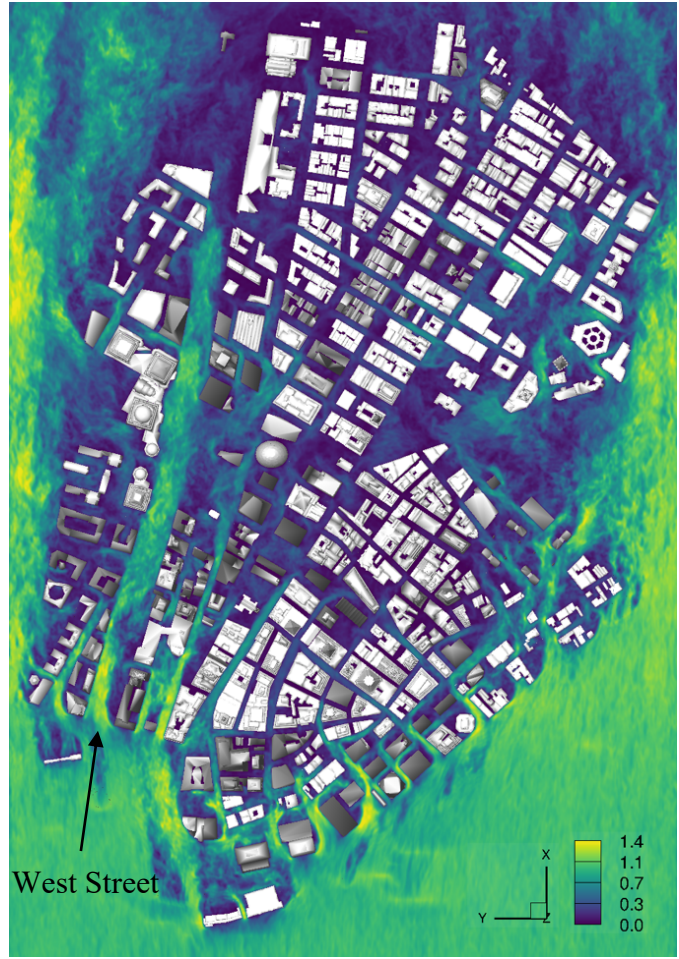

**Figure 12.** Contours of LES-computed instantaneous velocity magnitude, non-dimensionalized with the mean-flow velocity of the dominant wind (=3.58 m/s), on the horizonal plane that is approximately 13.4 m above the ground level. Buildings and ground are shown in grey. The wind flow is from the south (bottom) to the north (top).

**5.2 Time-averaged flow field**

The time-averaged flow field results of the wind flow field were obtained by averaging the instantaneous LES-computed flow field data for two flow-through periods. One flow-through time period is the time that it takes for a fluid particle to travel from the inlet to the outlet of the study area. Given the length of the study area in the windwise direction (~2.5 km) and the mean-flow velocity of the dominant wind (=3.58 m/s), one flow-through of the study area is about 11.7 min.

    Figure 13 depicts the time-averaged velocity magnitude, non-dimensionalized with the mean-flow velocity, on the three cross-planes "P1" to "P3". As seen, the time-averaged results show the formation of the shear-layer above the urban canopy. Low-momentum wind flow regions

can also be clearly seen in the areas deep into the city. More specifically, near zero wind flow velocity can be seen at low elevations and near ground. It is expected that the convective forces of the wind flow at near ground elevations to be insignificant. Therefore, we expect that the contaminant concentrations to have longer resident time in these low momentum regions of the city.

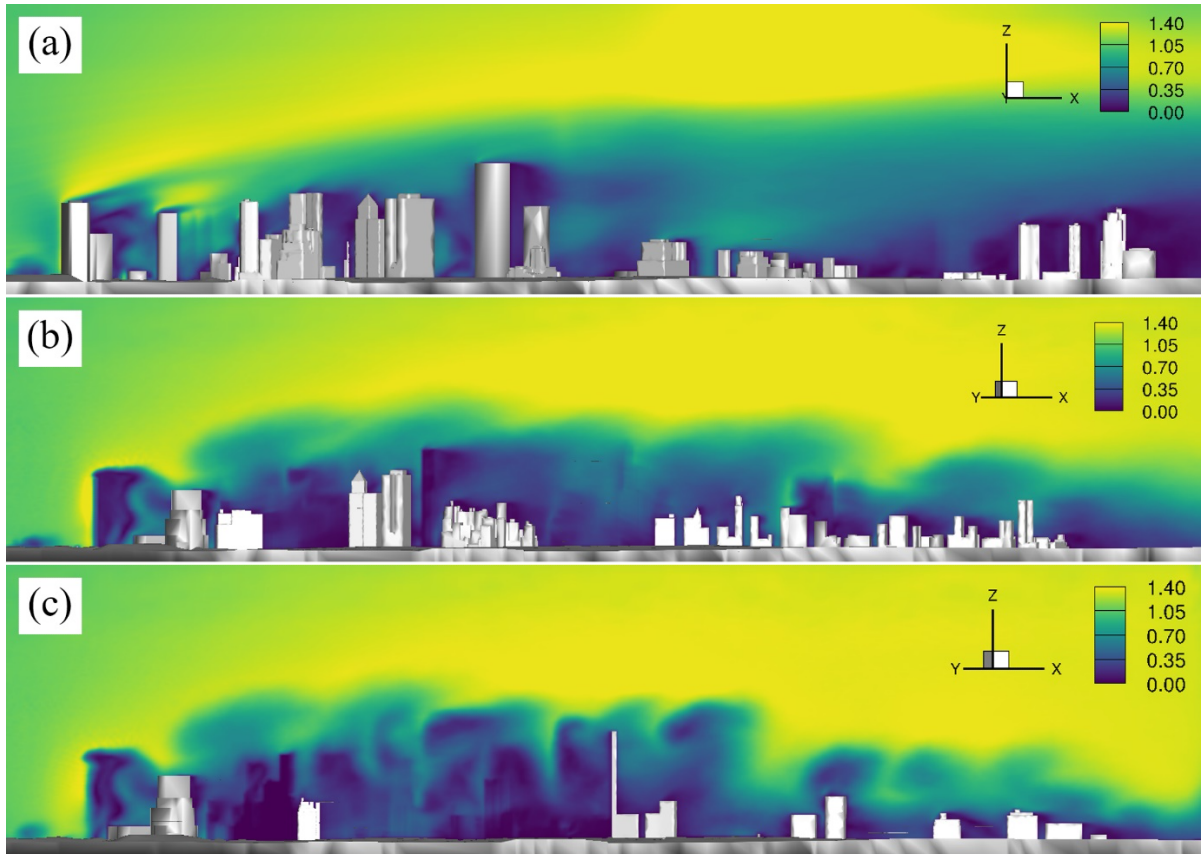

**Figure 13.** Contours of time-averaged velocity magnitude, non-dimensionalized with the mean-flow velocity of the dominant wind (=3.58 m/s), on the cross planes "P1" (a), "P2" (b), and "P3" (b). Location of the cross planes is shown in Fig. 10. Buildings and ground are shown in grey. The tallest building in (a) is the One World Trade Center. The wind flow is from the south (left) to the north (right).

The low-momentum regions of in the city can be clearly seen from the plan view of the time-averaged dimensionless velocity magnitude in Fig. 14, which shows the wind velocity field on the horizontal plane, which is about 13.4 m above the ground. As seen in this figure, channeling of the wind flow in the streets aligned in the windwise direction has led to high velocity in some streets (such as West Street). While the shielding effect of the buildings has induced near zero momentum in their lee side.

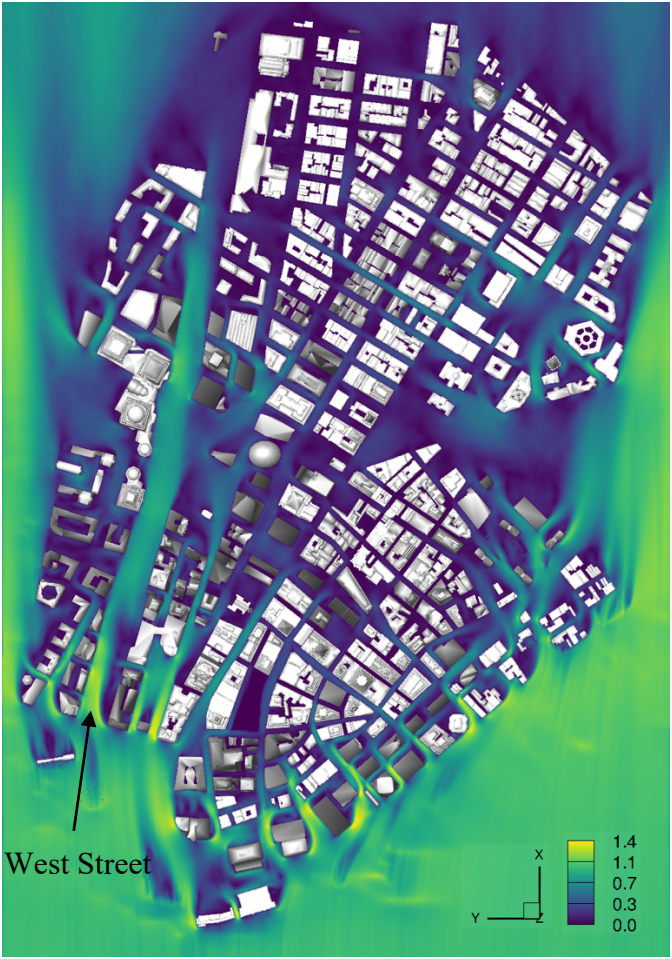

**Figure 14.** Contours of time-averaged velocity magnitude, non-dimensionalized with the mean-flow velocity of the dominant wind (=3.58 m/s), on the horizonal plane that is approximately 13.4 m above the ground level. Buildings and ground are shown in grey. The wind flow is from the south (bottom) to the north (top).

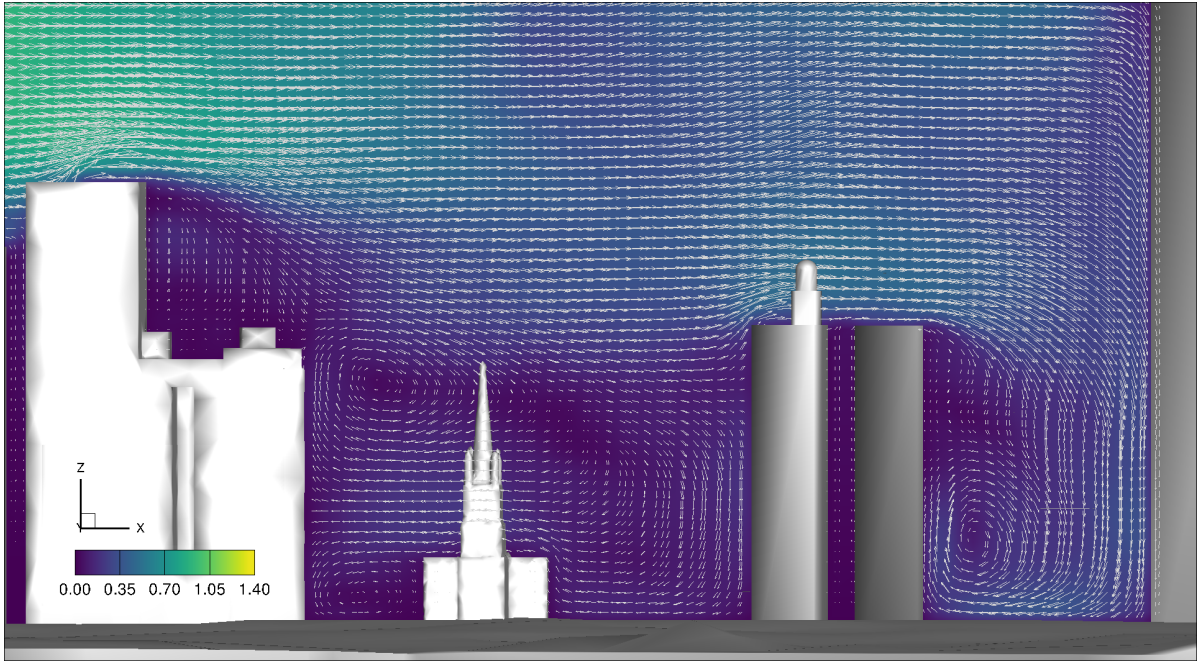

**Figure 15.** Contours of time-averaged velocity magnitude, non-dimensionalized with the mean-flow velocity of the dominant wind (=3.58 m/s), superimposed with velocity vectors on the cross plane "P3". Flow is from the south (left) to the north (right).

To better illustrate the complex flow patterns around buildings, we plot in Fig. 15 the contours of time-averaged dimensionless velocity magnitude, superimposed by velocity vectors, around four buildings on the cross plane "P3". As seen, downdraft wind flows occurs at the windward face of buildings forming recirculation zone near street level. In the leeside of buildings, however, updraft wind flow patterns take place. The combination of these dynamics leads to diverging/converging flow patterns at street level and extends for several blocks. For example, one can clearly observe such downdraft and updraft flow patterns around One Chase Manhattan Plaza (see Fig. 16) and One World Trade Center (see Fig. 17). In these figures, for the sake of clarity, we plot the time-averaged streamlines in a limited portion of the flow field. The time-averaged vertical velocity component (non-dimensionalized with the mean-flow velocity of 3.58 m/s) on the windward face of One Chase Manhattan Plaza is about -0.25. This marks the significance of the downdraft wind flow on the windward face of the building. This same effect can also be seen on the windward side of One World Trade Center. The updraft flow pattern on the leeside of the buildings can be clearly seen in Fig. 16 and 17, as the time-averaged streamlines are extended upward and away from ground leveling off at the peak of the buildings. The time-averaged dimensionless vertical velocity component in the leeside of the One World Trade Center is about 0.1 marking the strength of the updraft flow in the leeside of the building.

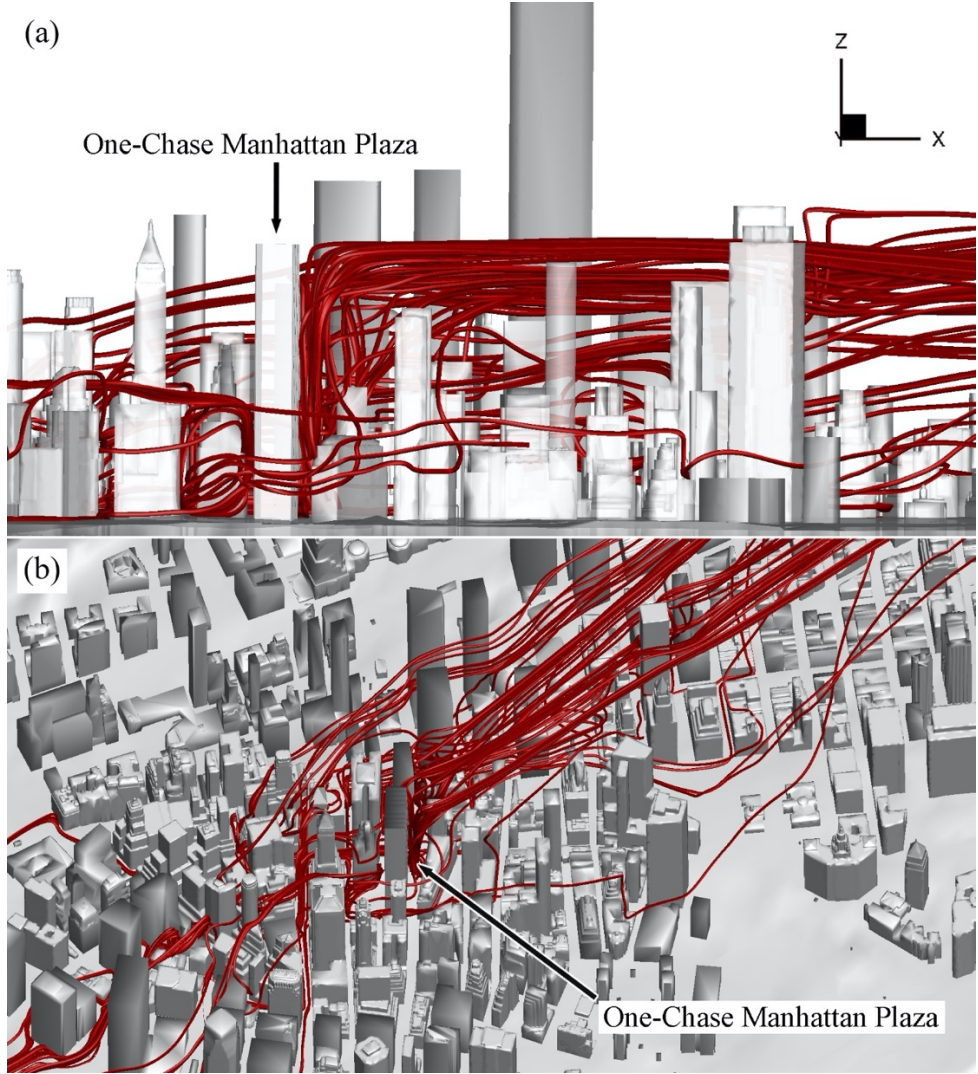

**Figure 16.** Time-averaged streamlines showing updraft flows on the leeside of several buildings including One Chase Manhattan Plaza. (a) shows a cross plane looking west with the flow from left to right. (b) shows a 3D view of the study area zoomed on the One Chase Manhattan Plaza in which the flow is from south (bottom left) to north (top right).

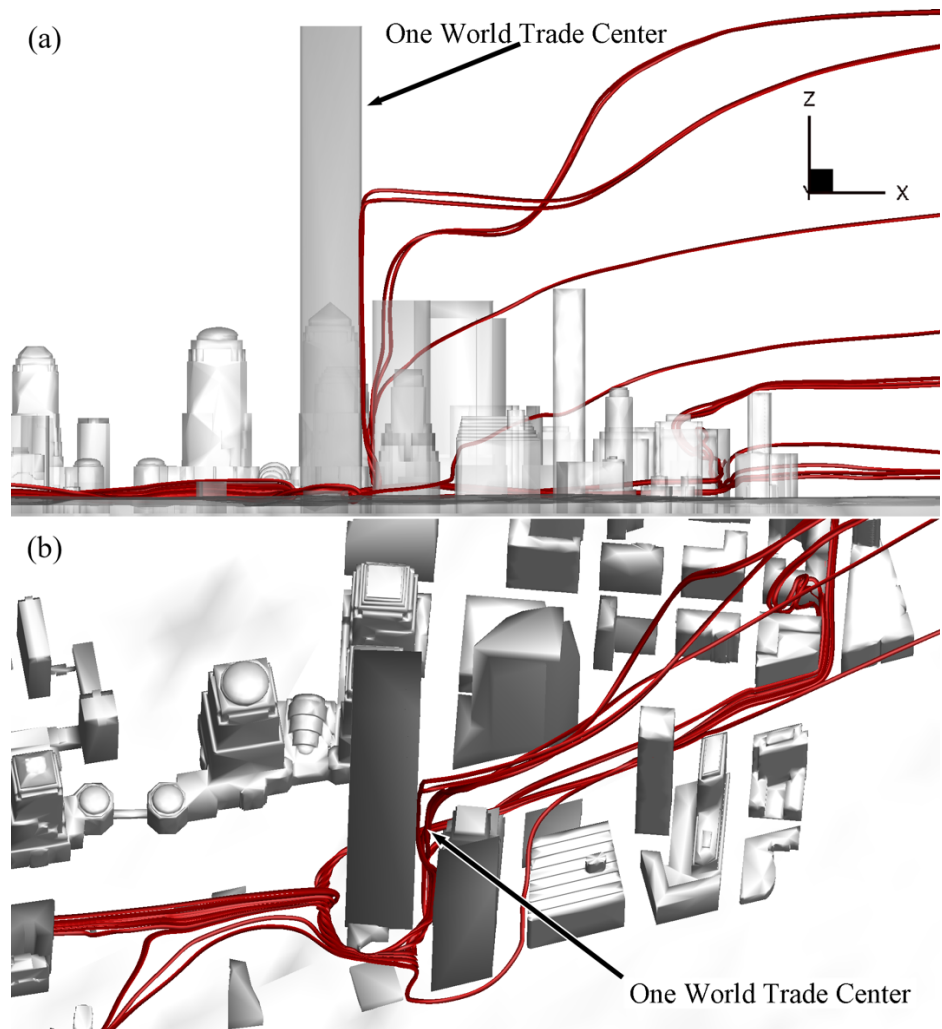

**Figure 17.** Time-averaged streamlines marking the updraft of the wind flow on the leeside of One World Trade Center. (a) shows a cross plane looking west with the flow from left to right. (b) shows a 3D view of the study area zoomed on the One Chase Manhattan Plaza in which the flow is from south (bottom left) to north (top right).

## 5.3 Contaminant concentration field

Once the wind flow field in of the study area was statistically converged, we continued the LES of the flow field for one more flow-through time (~11.7 min), before activating the convection-diffusion module of the model to simulate the spreading of the contaminant concentration. This allowed us to have a realistic fully turbulent wind flow field in the study area. Then, we started the point-source release of the contaminant allowing for the plume of contamination to spread through Lower Manhattan. As illustrated in Fig. 9, the point-source sphere is located on the east side of the NY Stock. A constant non-dimensional concentration value of 1.0 (volume fraction) was applied at the source-point sphere of contaminant using a Dirichlet boundary condition for 19.1 min, when the contamination plume transported and reached the end of the study domain. We denote this period of plume spreading as the ascending phase of plume transport.

After 27.4 min, the contamination plume exited the study domain. At this time, we stopped feeding the contaminant concentration at the source point and continued the simulation to allow for the residual of contamination plume to exit the study area via the turbulent wind flow. This period of plume dispersion is denoted as descending phase of plume transport. In this section, we first present the results of plume transport during the ascending phase of transport. Then, the simulation results of the contaminant transport during the descending phase of transport are presented.

**5.3.1 Contaminant transport during ascending phase**

Soon after the release of the contaminant from the point-source, a plume of high contaminant concentration forms. The transport process, which is governed by the diffusion and convection of the turbulent wind flow, during the ascending phase can be seen in Fig. 18 in which we plot snapshots of the plume concentration at various instances in time after the pollution release on the vertical cross-plane "P1". For a time sequence of the movement of the contaminant, a complete plume transport process can be seen in Movie 1 which is created using the instantaneous results of the plume evolution during the ascending phase of transport on cross-plane "P1". As seen in Fig 18, shortly after their release of the contaminant, the concentration spreads in the windwise direction reaching the end of the study area, i.e. the north of Lower Manhattan, after about 19.1 min. As seen in Fig.18, several block downwind of the point-source of contamination, the plume rises well above the ground and street level owing to the updraft wind flow pattern. Therefore, this demonstrates that the updraft wind flow pattern, which was discussed in Section 5.2, does have important implications for the contaminant plume transport. More specifically, immediately downwind of the point-source, the contaminant plume deems to be blocked by the row of buildings (i.e., Wall Street and the Equitable Life) and, consequently, is lifted up by the updraft wind flow. Despite the quick rise of the contaminant plume, its vertical rise is strictly limited to the urban canopy as the contaminant particles do not cross the shear layer (Fig. 18(e) and Movie 1).

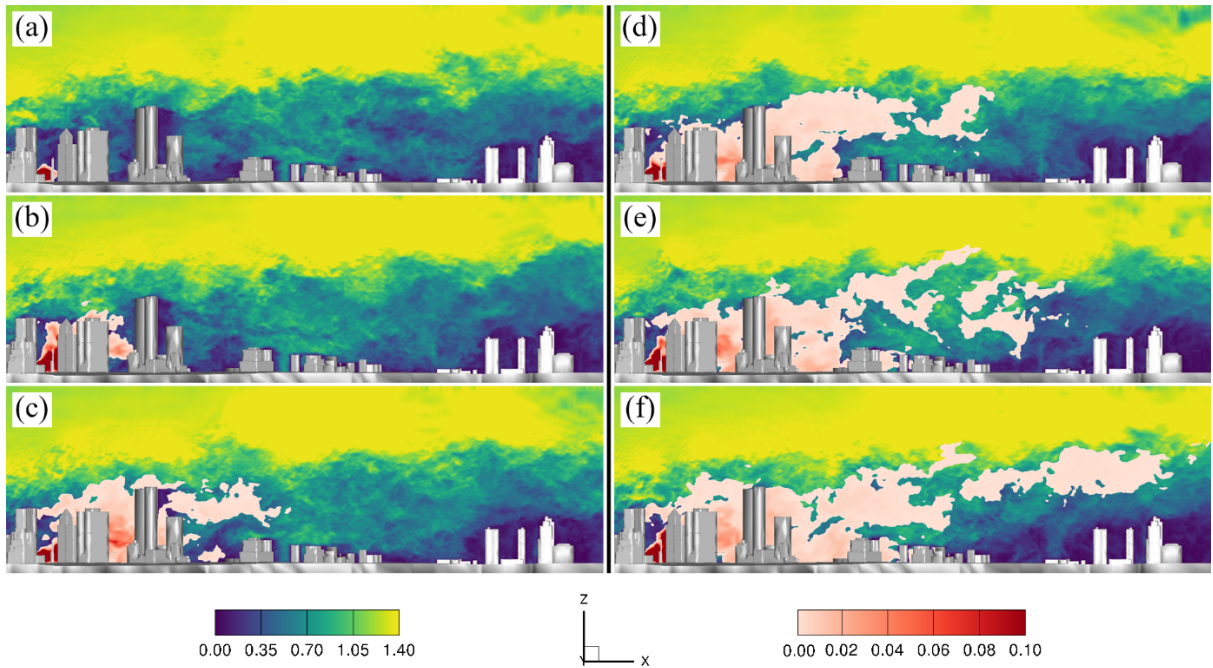

**Figure 18.** Contours of instantaneous simulation results on the cross-plane "P1" during the ascending phase of contaminant concentration transport marking the vertical spreading of the contaminant plume. Contours of contaminant concentration (in red range) are superimposed over the contours of dimensionless velocity magnitude. (a), (b), (c), (d), (e), and (f) correspond to times t = 28 s, 4.2 min, 8.3 min, 12.5 min, 6.8 min, and 20.0 min, respectively. The concentration less than 1 part per thousand are not shown. The flow is from the south (left) to the north (right).

To examine the horizontal transport of the contaminant plume and its spanwise expansion, we plot in Fig.19 iso-surfaces of the contaminant concentration from the top view. In Movie 2, we also show the evolution of the iso-surfaces of contaminant concentration (of 0.1 in blue and 0.2 in red) from the top view. As seen in this figure and the animation video, the width of the area affected by the contaminant plume increases quickly downwind of the release point reaching several blocks of Lower Manhattan in the spanwise direction. It can be clearly seen in Movie 2 that the contaminant plume meanders in horizontal direction and exposes the city block on the edge of the plume to fluctuating amounts of contaminant concentrations. The meandering effect becomes more significant after about 15 min, when the plume reaches the end of the study area in Lower Manhattan (Fig 19(a)).

Starting at t = 19.1 min, when the plume reaches the end of the flow domain, the concentration field was time-averaged for one flow-through time (~11.7 min). The time-averaged concentration field on the vertical cross-plane "P1" is shown in Fig. 20. As seen in this figure, the contaminant plume extends all the way to the outlet of the flow domain. The leeside of skyscrapers seems to trap high concentrations of contaminants in their recirculation zone. In addition, at about 1 km downwind of the source-point, the contaminant plume rises up creating a near zero contaminant region at the street level. The rise of the contaminant plume is expected to occur due to the updraft wind flow pattern deep into the urban canopy (as mentioned in Section 5.2). Furthermore, the side expansion of the contaminant plume can be evaluated using the time-averaged iso-surfaces of contaminant concentration (see Fig. 21). As seen, a few blocks downwind of the source-point, the contaminant plume reaches a width of about 0.5 km. This spanwise expansion stays roughly the same throughout the flow domain.

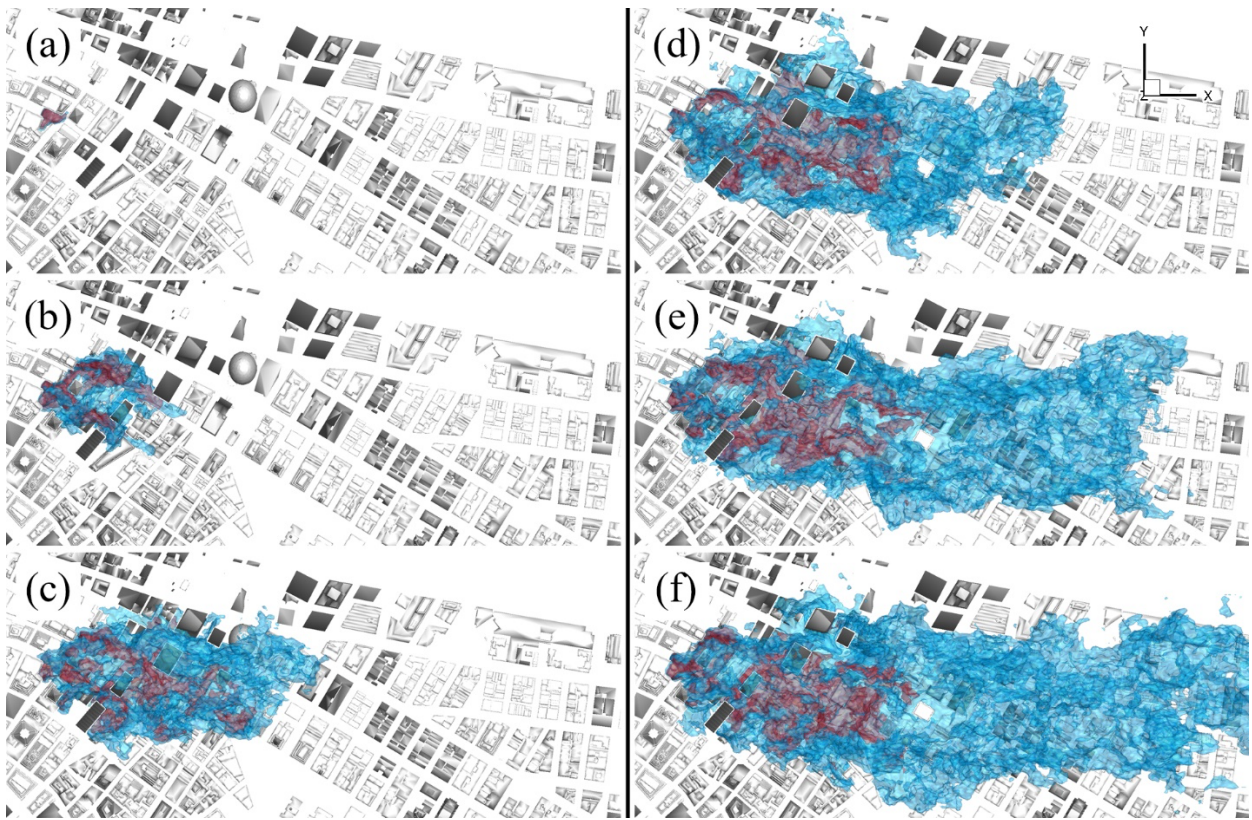

**Figure 19.** Iso-surfaces of instantaneous contaminant concentration from the top view. The blue and red iso-surfaces correspond to concentrations of 0.001 and 0.01, respectively. (a), (b), (c), (d), (e), and (f) correspond to times t = 0.49 min, 4.2 min, 8.3 min, 12.5 min, 16.8 min, and 20.0 min, respectively. The flow is from the south (left) to the north (right).

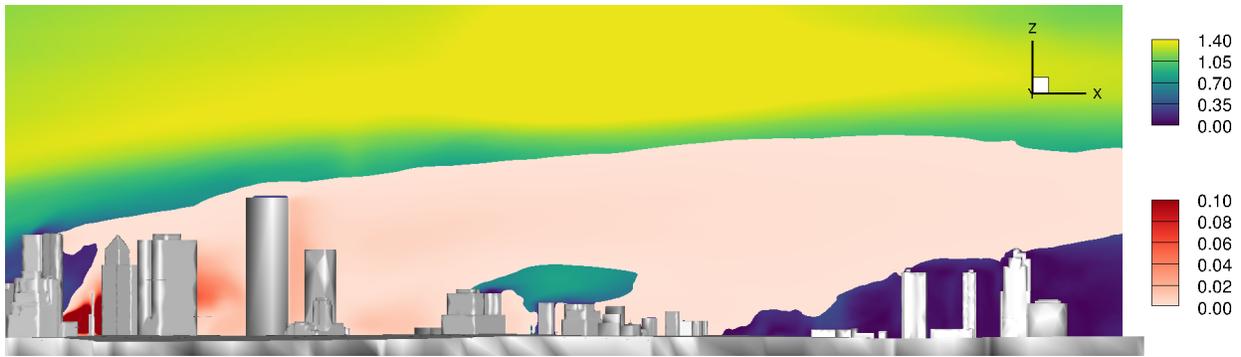

**Figure 20.** Contours of the time-averaged results on the cross-plane "P1" during the ascending phase of contaminant concentration transport marking the vertical spreading of the contaminant plume. Contours of contaminant concentration (in red range) are superimposed over the contours of dimensionless velocity magnitude. The concentration less than 1 part per thousand are not shown. The flow is from the south (left) to the north (right).

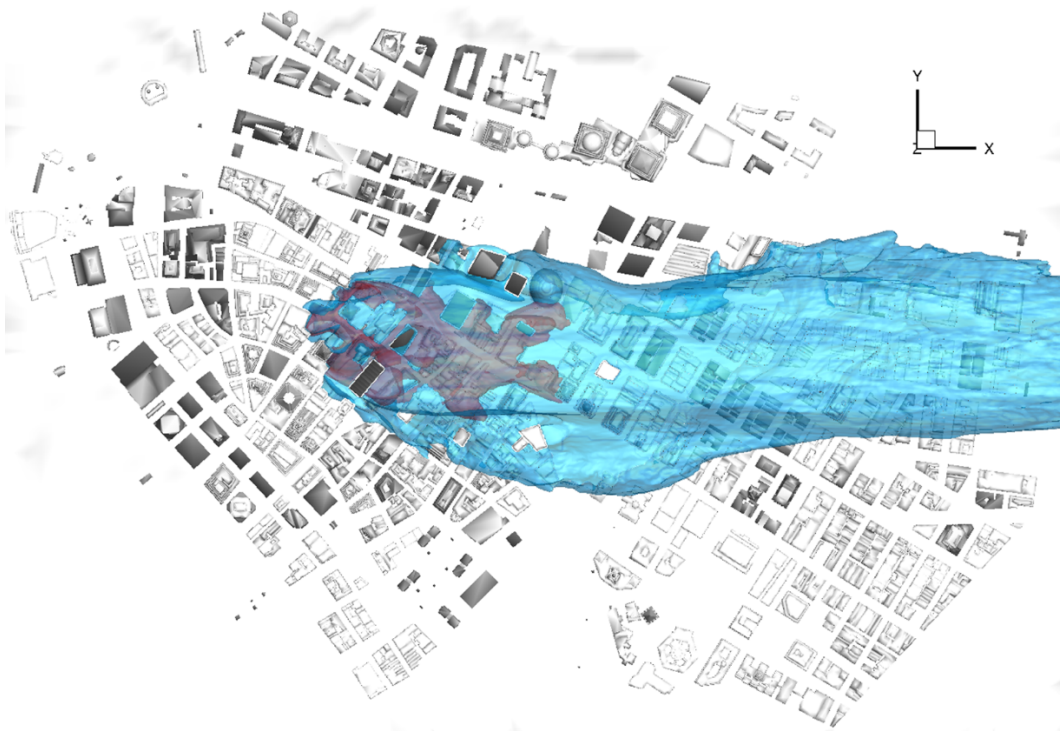

**Figure 21.** Iso-surfaces of the time-averaged contaminant concentration from the top view. The blue and red iso-surfaces correspond to concentrations of 0.001 and 0.01, respectively. The flow is from the south (left) to the north (right).

### 5.3.2 Contaminant transport during descending phase

The removal of the source of contamination is expected to lead to a gradual disappearance of the contaminant plume from the urban area. To investigate this process, we simulated a scenario in which the point-source of the contamination is removed. To do so, we used the concentration source at instant t = 46 min as the initial condition for a new simulation in which the source-point of pollution does not exit. Because of the lack of a source to feed the pollution into the flow domain, the overall concentration of contaminant in the urban area starts deceasing, i.e., the descending phase of concentration field begins. For the sake of discussion, in this section, we denote the start time of the simulation in the descending phase as t = 0.

Figures 22 and 23 depict the instantaneous contaminant concentration field at various times after the source-point of contaminant is removed. The receding process of the contaminant concentration can be clearly seen in the vertical cross-plane "P1" (Fig. 22) and a horizontal plane located 13.4 m above the mean ground level (Fig. 23). Also, we show in Movie 3 the evolution of contaminant concentration field during the descending phase of the pollution plume on cross-plane "P1". After roughly 39.4 min of time, the study area was cleared of contaminant concentration. Comparing the simulation results of Figs. 22 and 23, one can see that the contaminant concentration at higher elevation exit the domain much faster that those near the ground level. Also, the majority of the contaminants (mostly those at higher elevations above the ground) are cleared out of the flow domain in about 20 min. Between 20 min and 39.4 min, only small pockets of contaminant concentration, which are trapped in the leeside of buildings, remain in the flow domain resisting the convective force of the wind flow.

During the ascending phase, the time required for the contaminant to travel from the point-source to the end of the flow domain was determined to be 19.1 min (see Section 5.3.1). However, the receding time of the pollution (39.4 min) is roughly two times longer than that of the ascending (19.1 m). The marked difference between the time periods illustrates the crucial rule of buildings and other cityscape features in delaying the complete removal of the contaminants from the urban area. Such delay in recession of contaminant concentration, which is known as resident time, is mainly caused by the recirculation zones in the leeside of the buildings where pockets of contaminant remain trapped for relatively long periods of time. Moreover, the contaminant concentration field is advected in the windwise direction and mainly through the street canyons that are more or less parallel to the dominant wind direction (Fig. 23).

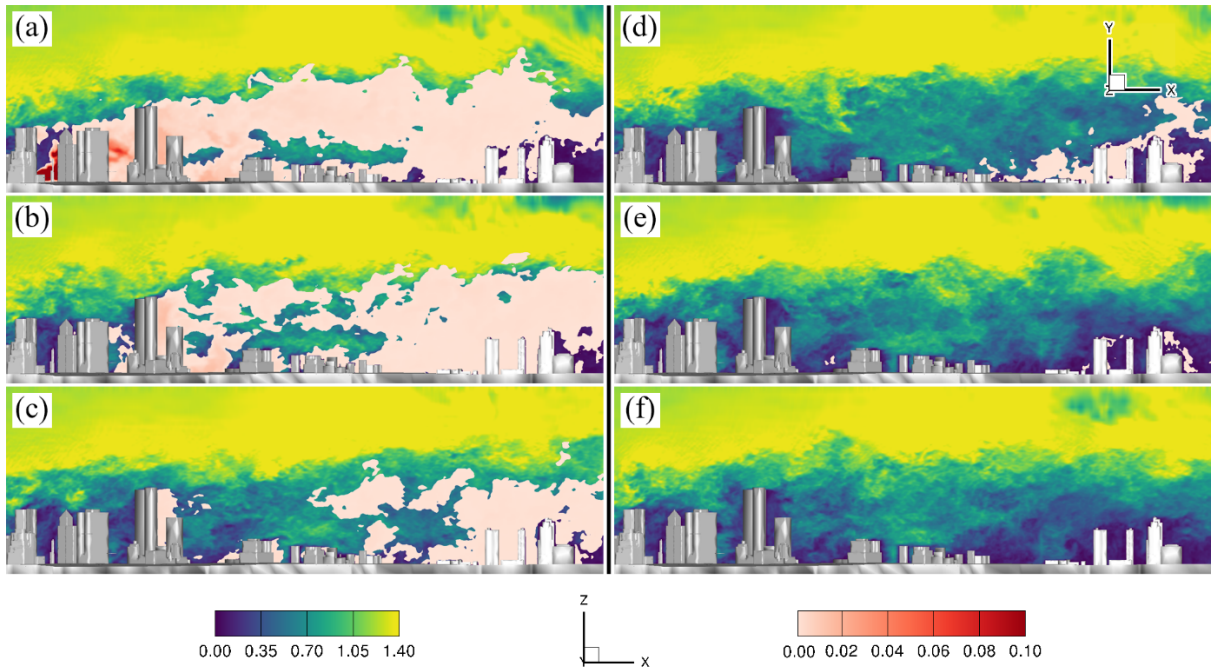

**Figure 22.** Contours of instantaneous simulation results on the cross-plane "P1" during the descending phase of contaminant concentration transport. Contours of contaminant concentration (in red range) are superimposed over the contours of dimensionless velocity magnitude. (a), (b), (c), (d), (e) and (f) correspond to times t = 0 min, 4.2 min, 8.4 min, 16.8 min, 33.6 min, and 46.7 min, respectively. The concentration less than 1 part per thousand are cut off. The flow is from the south (left) to the north (right).

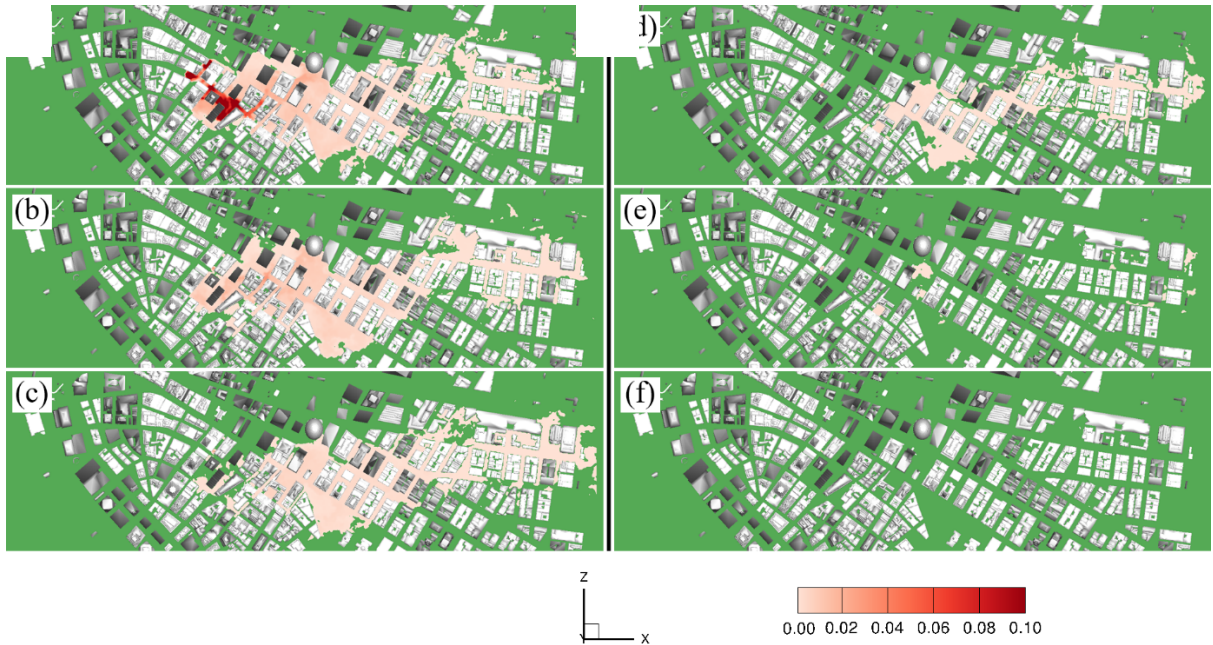

**Figure 23.** Contours of instantaneous simulation results on the horizontal plane 13.4 m above the mean ground level during the descending phase of contaminant concentration transport. (a), (b), (c), (d), (e) and (f) correspond to times t = 0 min, 4.2 min, 8.4 min, 16.8 min, 33.6 min, and 46.7 min, respectively. The timing is calculated since the removal of the source-point of contamination. The concentration less than 1 part per thousand are cut off. The flow is from the south (left) to the north (right).

## 6 Conclusion

We carried out LES of a micro-scale model of a large section of New York City under a dominant wind conduction (i.e., south to north wind with a mean-flow velocity of 3.58 m/s) to investigate the spreading of contaminant concentration from a source-point in Lower Manhattan. The computational grid used in this study is significantly finer than previously reported simulations to study wind flow and contaminant transport in urban areas.

As discussed in Section 5.3, shortly after the release of the contaminants (i.e., about 19 min), the pollution concentration spreads several blocks in spanwise direction covering a width of about 0.5 km in the spanwise direction. At this point, the pollution plume extends all the way from the point-source to the outlet of the domain along the windwise direction. The spreading of pollution in the vertical direction is strictly limited to the shear layer of the urban canopy. During this phase of pollution transport, which is denoted as the ascending phase, the concentration of contaminants is highest near the source point and also in the leeside of the large and small building at elevations near the ground and street level. When the source of pollution is removed, the contaminant concentration starts receding from the urban area. The receding process, which we denoted as descending phase of transport, takes about 39.4 min, for the pollution to be cleared from the study area. The recirculation zones in the leeside of buildings contributed to the relatively longer resident time of the pollution in the urban area. Our findings regarding the spreading and recession processes of the contaminant in such a highly populated urban area can be of importance for strategic planning by decision makers and stakeholders.

Finally, we should note that because of the high computational cost of the simulations, we limited our study to a unidirectional wind condition. Our numerical method, however, can be used to computationally investigate a variety of ambient flow conditions as well as other source-point locations.

**Acknowledgments**

This work was supported by a sub-award from the National Institute of Health (2R44ES025070-02). We wish to thank Filippo Coletti and Alec Petersen of ETH Zürich for providing us with their experimental data to validate the numerical model.

**Data Availability Statement**

The data that support the findings of this study are available from the corresponding author upon reasonable request.